\documentclass[letterpaper,twocolumn,fleqn]{article} 

\usepackage{ist}
\usepackage{amsmath}
\usepackage[ruled,vlined]{algorithm2e}
\usepackage{array}
\usepackage{multirow}
\usepackage{tabularx}

\title{Proximal Newton Methods for X-Ray Imaging with Non-Smooth Regularization}

\author{Tao Ge, Umberto Villa, Ulugbek S. Kamilov, Joseph A. O’Sullivan ;\\
Mckelvey School of Enginerring, Washington University in St. Louis, Saint Louis, MO}

\date{} 

\begin{document} 

\maketitle 

\thispagestyle{empty} 


\begin{abstract}
Non-smooth regularization is widely used in image reconstruction to eliminate the noise while preserving subtle image structures. In this work, we investigate the use of proximal Newton (PN) method to solve an optimization problem with a smooth data-fidelity term and total variation (TV) regularization arising from image reconstruction applications. Specifically, we consider a nonlinear Poisson-modeled single-energy X-ray computed tomography reconstruction problem with the data-fidelity term given by the I-divergence. The PN algorithm is compared to state-of-the-art first-order proximal algorithms, such as the well-established fast iterative shrinkage and thresholding algorithm (FISTA), both in terms of number of iterations and time to solutions. We discuss the key factors that influence the performance of PN, including the strength of regularization, the stopping criterion for both sub-problem and main-problem, and the use of exact or approximated Hessian operators. 
\end{abstract}

\section{Introduction}
\label{sec:intro}

Structures in medical images are critical, providing information inside the human body for diagnosis and development of treatment plans. However, an improper regularization term may eliminate subtle structures during image reconstruction \cite{Charbonnier_1997, Hao_2018}. Non-smooth regularization terms, such as $\ell_1$ norm or total variation (TV) \cite{RUDIN1992259, Dobson_1996}, are widely used in image reconstruction to enforce sparsity (either in the image, wavelet, or gradient domain) while reducing noise. For these methods to be useful in clinical diagnosis and research, the images should be reconstructed within a clinically acceptable time. 

The use of TV regularization is often advocated in imaging applications because of its ability to preserve sharp edges in the reconstructed image; however, the non-smoothness of the TV functional poses a substantial challenge to the efficient solution of the corresponding optimization problem.

Proximal gradient algorithms (such as FISTA) \cite{Beck_2009_FISTA} represent the state-of-art to solve image reconstruction problems with non-smooth regularization terms like TV. However, these methods still require hundreds of iterations to converge and each iteration requires to evaluate the smooth data fidelity term and its gradient. This can be extremely time-consuming for problems, such as transmission tomography, where the imaging operator is expensive to compute \cite{Lange_1984, O_2006}.

Our goal is to investigate more efficient solution algorithms for this class of problems that exploit curvature information (the Hessian of the smooth data-fidelity term) to achieve faster convergence than classical proximal gradient methods. Specifically, we numerically show that, for our target application, Proximal Newton method can outperform state-of-the-art proximal gradient methods, such as FISTA, and we also discuss key factors that influence the performance of the algorithm.

Second-order proximal methods, and proximal Newton in particular, have mostly been explored in the fields of bioinformatics \cite{Zhang_2018}, signal processing \cite{Antonello_2018}, and statistical learning \cite{Liu_2017, Yuchen_2015}. To the best of our knowledge, this is the first time that proximal Newton methods are applied to an image reconstruction problem with TV penalty term. Our results suggest that for our target problem, and possibly for many other problems in which evaluations of the smooth data fidelity term are computationally expensive, PN is faster than state-of-the-art proximal first order methods, such as FISTA.

\section{Method}
We consider the problem of minimizing a composite function 
\begin{equation}
    f(x)=l(x)+h(x), 
\end{equation}
where $l(x)$ is the smooth data fidelity term and $h(x)$ is the non-smooth penalty term.
The proximal Newton method minimizes the composite function $f(x)$  by successively constructing and minimizing a surrogate function $\tilde f_k(x)$ \cite{Vishwanathan_1997, Jason_2014}.  At each iteration $k$, the PN algorithm performs three steps.\\
\textbf{1. Surrogate function construction.} 

In the first step, the PN method computes the surrogate function $\tilde f_k(x)$ by approximating the smooth data fidelity term $l$ with its second order Taylor approximation centered at the current iterate $x_k$, that is 
\begin{multline}\label{eq:surrogate}
    \tilde f_k(y)=l(x_k )+\nabla l(x_k )^T (y-x_k )\\
    +\frac{1}{2} (y-x_k )^T H_k (y-x_k )+h(y),
\end{multline}
where $\nabla l(x_k)$ and $H_k$ are the gradient and Hessian of function $l(x)$ at $x_k$. \\
\textbf{2. Surrogate minimization.} 

In the second step, the PN method computes a search direction $d_k= y-x_k$ by minimizing $\tilde f(y)$, that is 
\begin{equation}
\label{eq:d}
    \hat d_k= \arg\min_{d_k}  \nabla l(x_k )^T d_k+\frac{1}{2} d_k^T H_k d_k+h(x_k+d_k ),
\end{equation}
This subproblem does not usually admit a closed-form solution and need to solved iteratively, possibly using a first-order proximal algorithm such as FISTA. Note that the solution of this subproblem does not require any evaluations of the (computationally expensive) data fidelity term or of its gradient.\\
\textbf{3. Solution update.} 

In the third and last step, the PN method updates the iterate $x_k$ as
\begin{equation}
    x_{k+1}=x_k+t d_k,
\end{equation}
where the step size $t \in (0,1]$ is calculated by back-tracking line search to ensure the sufficient descent condition 
\begin{equation}
\label{eq:suff_decend}
    f(x_k+t d_k )\leq f(x_k )+at(\nabla l(x_k )^T d_k+h(x_k+d_k )-h(x_k ))
\end{equation}
for some $a \in (0,0.5]$.

The pseudo-code for PN algorithm is shown in Algorithm \ref{alg:1}. 

\begin{algorithm}
\SetAlgoLined
\KwResult{$x$ }
 Initialization: $x_0=0$\;
 \While{Not Converge}{
  Evaluate the gradient $\nabla l(x_k)$ and Hessian $H_k$ of $f(x)$ at $x=x_k$;\\[0.5mm]
  Use a first-order proximal gradient method to solve 
  $\hat{d}_k=\arg\min_{d_k}\nabla l(x_k)^Td_k+\frac{1}{2}d_k^TH_kd_k+h(x_k+d_k);$
  Set $t=1$;\\
  \While{Sufficient descent condition \eqref{eq:suff_decend} not satisfied}{
   Set $t=0.7\times t$;
   }
   Update $x_{k+1}=x_k+t\hat{d}_k$
 }
 \caption{Proximal Newton-Type Method}
 \label{alg:1}
\end{algorithm}

It is worth to notice that the performance and practical usefulness of the PN method strongly depends on the availability of computational efficient solver for the surrogate minimization problem in \eqref{eq:d}. To this aim, we consider the two complementary approaches proposed in \cite{Jason_2014}.

First, one can reduce the computational cost of evaluating the surrogate function $\tilde f$ and its proximal gradient by replacing the true Hessian matrix $H_k$ with a cheaper-to-apply Quasi-Newton approximation of $H_k$, such as that given by the (limited-memory) Broyden-Fletcher-Goldfarb-Shanno (BFGS) algorithm \cite{Nocedal_1980}.

Second, one can solve subproblem \eqref{eq:d} inexactly by an using early stopping criterion, such as a fixed maximum iterations or an adaptive stopping condition, to get an inexact solution with much less inner iterations. Simply put, one need not solve subproblem \eqref{eq:d} accurately when the surrogate function $\tilde f$ in \eqref{eq:surrogate} is an inaccurate approximation of the objective function $f$. Rather, to preserve the fast convergence rate of PN methods, one should solve \eqref{eq:d} accurately only when $x_k$ is near the optimal solution or $\tilde f$ is an accurate approximation to $f$. Following \cite{Jason_2014}, we then generalize the adaptive stopping condition originally introduced by Eisenstat and Walker in 1996 \cite{Eisensatat_1996} for smooth problems as follows:
\begin{multline}\label{eq:inexact_stopping}
    \left\|x_{k+1}-P^h\left(x_{k+1}-\nabla \tilde l_k(x_{k+1})\right)\right\| \\ \leq \eta_k \left\|x_k-P^h\left(x_k-\nabla l(x_k)\right)\right\|,
\end{multline}
where $k$ is the outer iteration index, and $P^h$ is the proximal mapping of the non-smooth term, and $\nabla \tilde l_{k}(x_{k+1})=H_{k}(x_{k+1}-x_{k})+\nabla l(x_{k})$ is the gradient of the approximated smooth term evaluated at $x=x_{k}$.
Here, $\eta$ is a forcing term that satisfies 
\begin{equation}
    \eta_k=\min \left\{0.1,\frac{\left\|P^h\left(x_k-\nabla l(x_k)\right)-P^h\left(x_k-\nabla\tilde l_{k-1}(x_k)\right)\right|}{\left\|x_{k-1}-P^h(x_{k-1}-\nabla l(x_{k-1}))\right\|}\right\}.
\end{equation}

\section{Target application}

In the numerical results presented in the next section, we consider the minimization of a composite function $f(x)$ arising in single-energy X-ray computed tomography (CT) reconstruction problems \cite{O_2006}.

Here, $x$ denotes the sought-after linear attenuation coefficient (image), the data fidelity $l(x)$ stems from a Poisson negative log-likelihood function, and the regularization term $h(x)$ is an isotropic TV function \cite{RUDIN1992259}
\begin{multline}
    h(x)=\lambda TV_I (x)\\
    =\lambda \sum_a\sum_b\sqrt{(x_{a,b}-x_{a+1,b } )^2+(x_{a,b}-x_{a,b+1}  )^2 },
\end{multline}
where $a,b$ denote a 2-D index in the image domain, and $\lambda $ controls the strength of regularization. The proximal mapping of total variation is solved by fast gradient projection \cite{Beck_2009}, which iteratively projects and updates the image in its gradient domain. 

In what follow we provide a derivation for the expression of the statistically inspired data fidelity term $l(x)$ and of its gradient and Hessian. The CT reconstruction problem is modeled as an estimation of means of independent Poisson random variables. The survival probability of a photon that is transmitted through the target follows the Poisson distribution, and the mean of the photon counts $g(x)$ is given by the Beer’s law:  
\begin{equation}
    g_j (x)=I_0 \cdot e^{-\sum_i a_{ij} x_i }=I_0 \cdot e^{-Ax }
\end{equation}
where  $i,j$ denotes the index in image (pixels) and measurement (ray) domain, respectively, $x$ is the attenuation coefficient to be estimated, $I_0$ denotes mean number of photons, and $a_{ij}$ represents the length of the path of ray $j$ in pixel $i$, whose matrix form is written as $A$. 

Maximizing the Poisson likelihood is equivalent to minimizing I-divergence function
\begin{equation}
    l(x)=\sum_j d_j \ln \frac{d_j}{g_j (x)} -d_j+g_j (x), 
\end{equation}
where $d$ is the measured transmission data obtained from the scanner. 
Finally, the gradient of $l(x)$ is given by 
\begin{equation}
\begin{split}
\nabla l(x)&=\frac{\partial}{\partial x}\left(\sum_j d_j\sum_i a_{ij}x_i+\sum_j e^{-\sum_ia_{ij}x_i} \right)\\
&=
\begin{bmatrix}
\sum_j d_j a_{1j}-\sum_j a_{1j}e^{-\sum_ia_{ij}x_i} \\ 
\vdots  \\
\sum_j d_j a_{Nj}-\sum_j a_{Nj} e^{-\sum_ia_{ij}x_i}
\end{bmatrix}\\
&=A^T(d-e^{-Ax}),\\
\end{split}
\end{equation}
and the second-order derivative of $l(x)$ reads
\begin{equation}
\begin{split}
&H(x)=\frac{\partial^2}{\partial x^2}\left(\sum_j d_j\sum_i a_{ij}x_i+\sum_j e^{-\sum_ia_{ij}x_i} \right)\\
&=
\begin{bmatrix}
\sum_j a_{1j}e^{-\sum_ia_{ij}x_i}A_{1j}    & \dots & \sum_j A_{1j}e^{-\sum_ia_{ij}x_i}a_{Nj} \\ 
\sum_j a_{2j}e^{-\sum_ia_{ij}x_i}A_{1j}    & \dots & \sum_j A_{2j}e^{-\sum_ia_{ij}x_i}a_{Nj} \\ 
\vdots  & \ddots & \vdots \\
\sum_j a_{Nj}e^{-\sum_ia_{ij}x_i}A_{1j}    &  \dots &  \sum_j A_{Nj}e^{-\sum_ia_{ij}x_i}a_{Nj}
\end{bmatrix}\\
&=A^TSA,
\end{split}
\end{equation}
where $S$ is a square diagonal matrix with entries $s_{ii}= e^{-(Ax)_i }$.

The matrix $H$ is formally a dense operator of size number of pixels by number of pixels and shall not be computed explicitly. 
For example, consider a two-dimensional reconstruction problem of a $256\times256$ image, then $H$ would be a $65526\times 65536$ dense matrix, those storage would require more than $34$ gigabytes memory using double precision. This memory requirement would then increase of a factor 16 when doubling the image resolution. To put things in perspective, a 3-dimensional CT reconstruction problem of practical relevance commonly involves images with resolution of $512\times 512\times 200$ pixels, thus requiring about 21 petabytes memory to store the Hessian matrix, which is impossible for the nowadays computational devices.

On the other hand, the PN algorithm does not require to explicitly access to the matrix $H$, but only the ability to compute the action of $H$ in a direction $d$. This action can be computed \emph{matrix-free} at the cost of performing one set of forward and back-projection steps as summarized in Algorithm \ref{alg:2}. It is worth noticing that if the forward and back-projection steps are also computed \emph{matrix-free}, then the only memory requirement is that of storing two vectors on the same dimension of the data, namely the scaling vector $s_k = e^{-(Ax_k)}$ and the intermediate vector $t$.

\begin{algorithm}
\SetAlgoLined
\KwIn{Direction $d$, Scaling vector $s_k = e^{-(Ax_k)}$ }
\KwResult{Hessian action $y = H_k d$ }
Forward projection step: $t = A d$;\\
Inplace Hadamard product: $t \leftarrow t \odot s_k$;\\
Backward projection step: $y = A^T t$.
 \caption{Matrix-free action of the Hessian $H_k$. The scaling vector $s_k = e^{-(Ax_k)}$ should be computed only once and stored.}
 \label{alg:2}
\end{algorithm}

\section{Result}

For all three sets of experiments presented below, we consider a CT system with 90 equispaced projection spanning a 180-degree angle and 90 rays per projection. The Shepp-Logan phantom generated by MATLAB built-in function is used as ground-truth. The noiseless measured data is obtained by exponentiation of the forward-projection of the ground truth, 
\begin{equation}
    d_{noiseless}=I_0 \cdot e^{-Ax_{true}},
\end{equation}
while the noisy data is a realization from a Poisson random vector whose mean is equal to the noiseless data
\begin{equation}
    d_{noisy}\sim Poisson(I_0 \cdot e^{-Ax_{true}}).
\end{equation}

In each experiment, the strength of regularization $\lambda$ is chosen by the generalized L-curve criterion. Unless otherwise indicated, default reconstruction parameters and PN algorithm settings are shown in the table below.

\begin{table}[ht]
\centering
\begin{tabularx}{0.4\textwidth} { 
  | >{\raggedright\arraybackslash}X 
  | >{\centering\arraybackslash}X 
  | >{\raggedleft\arraybackslash}X | }
 \hline
 Image Size & $64\times 64$ pixels\\
 \hline
 Sinogram Size & $90\times 90$ pixels\\
 \hline
 Diameter of FOV & 512 mm \\
 \hline
 $I_0$ & $1\times 10^5$ \\
 \hline
 $\lambda$ & $1\times 10^{-4}$ \\
 \hline
 $\alpha$ & $0.5$\\
 \hline
 Hessian Computation & Exact\\
 \hline
 Initial Guess & Zeros\\
  \hline
 \multirow{3}{4em}{PN Stopping Criterion}  & $\frac{|f(x_k)-f(x_{k-1})|}{|f(x_{k-1})|} \leq 10^{-4}$\\
 & $\frac{||x_{k'}-x_{k'-1}||}{||x_{k'-1}||} \leq 10^{-5}$\\
 & Max iter: 500\\
 \hline
  \multirow{4}{4em}{Subproblem Stopping Criterion}  & $\frac{||x_k-x_{k-1}||}{||x_{k-1}||} \leq 10^{-8}$\\
 & Max iter: 500\\
 & Adaptive Stopping Condition\\
\hline
\end{tabularx}
\label{tab: Default settings}
\end{table}

\begin{figure}[!hb]
  \includegraphics[width=1\columnwidth]{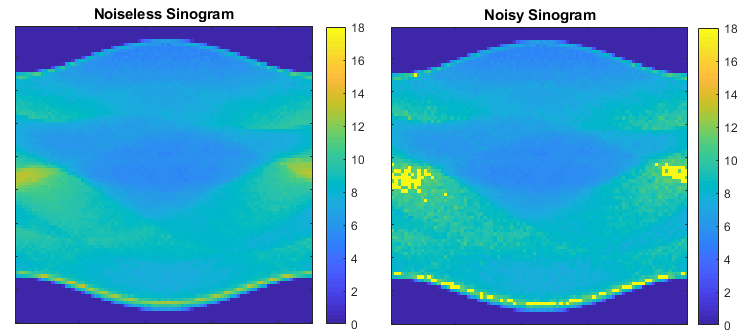}
  \caption{Simulated noiseless and noisy sinogram in negative log-scale}
  \label{Figure:sinogram}
\end{figure}

Figure \ref{Figure:sinogram} shows the negative log-scaled simulated transmission data with and without noise. It is worth noticing that we have clipped the range of values in the log-scaled sinogram for visualization purposes. Extremely bright-yellow values in the noisy sinogram correspond to detectors that fail to receive any photon.

For the computation below, we use a modified version of the PNOPT code \cite{Jason_2014} and TFOCS package \cite{Becker_2011} to solve the optimization problem using PN and FISTA, Airtools \cite{HANSEN20122167} to compute forward projection operator $A$, and TV proximal function in UNLocBoX \cite{PerraudinSPV14}. The forward projection operator $A$ was precomputed before solving the optimization problem and stored in a compressed sparse column (CSC) format to allow for fast application of the forward and back-projection steps within MATLAB.

Finally, the computer utilized for time analysis has a 20-threaded Intel Xeon E5 2630-v4 and 128 Gigabytes memory.
No GPU computation is involved.

\subsection{Efficiency comparison between PN and FISTA}

\begin{figure}[!hb]
  \includegraphics[width=1\columnwidth]{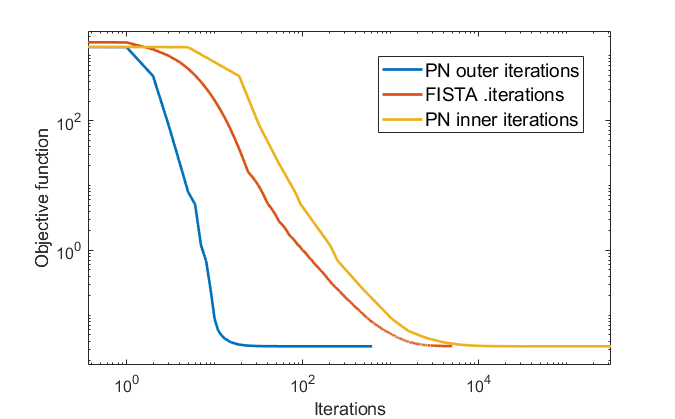}
  \caption{Objective function vs. number of iterations of PN and FISTA}
  \label{Figure:iter}
\end{figure}

The convergence rate of proximal Newton is compared to FISTA for an image size of $64\times 64$ pixels. Due to the higher convergence rate of second-order methods, PN greatly outperforms FISTA algorithm with respect
to the number of outer iterations, as shown in Figure \ref{Figure:iter}. PN takes approximately 20 iterations to achieve the minimum objective value, while FISTA takes more than 2,000 iterations. However, each PN outer iteration is more expensive that one FISTA iteration, since it requires solving subproblem \eqref{eq:d} iteratively. In total, the cumulative number of PN inner iterations is about five time larger than the number of FISTA iterations. 

\begin{figure}[!hb]
  \includegraphics[width=1\columnwidth]{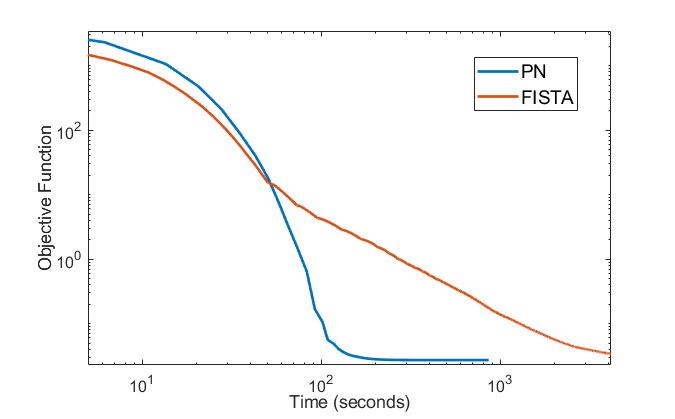}
  \caption{Objective function vs. time for PN and FISTA}
  \label{Figure:time}
\end{figure}

It is worth to notice that PN inner iterations are performed using the surrogate function $\tilde f$ in \eqref{eq:surrogate}, and therefore are computationally less expensive then FISTA iterations.
For this reason, Figure \ref{Figure:time} compares the two algorithms in terms of computational time. Overall, PN achieves a 3$\times$ speedup in terms of time-to-solution compared to FISTA.

Comparing Figure \ref{Figure:iter} and Figure \ref{Figure:time}, we can argue that the computational cost of each inner iteration in the PN method is extremely less than the computational cost of each iteration in FISTA. In fact, PN only computes gradient and Hessian of the data-fidelity term in the outer iteration, while FISTA computes the gradient in every iteration. Thus, for problems in which evaluations of the smooth data fidelity term are computationally expensive, PN is more efficient.

\begin{figure*}[!ht]
  \includegraphics[width=\textwidth]{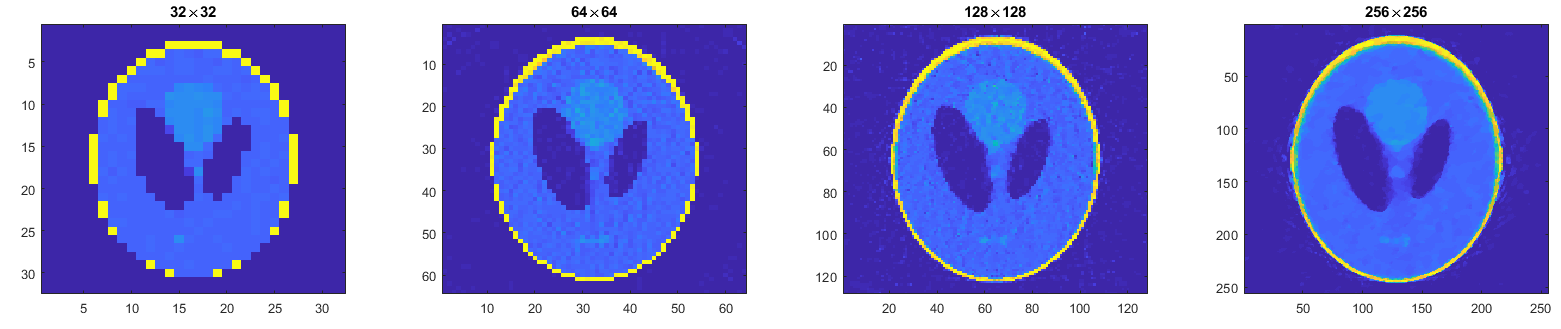}
  \caption{Reconstructed images with the resolutions of $32\times32$, $64\times64$, $128\times128$ and $256\times256$}
  \label{Figure:images}
\end{figure*}

\subsection{Scalability of PN with respect to image resolution }
Figure \ref{Figure:images} shows the reconstructed solution of the CT problem using the PN algorithm for different image resolutions. The image size ranges from $32\times32$ to $256\times256$ pixels with increases by a factor of $2\times 2$, corresponding to the pixel size ranging from $16\times16$ $ mm^2$ to $2\times2$ $ mm^2$, respectively. Also, the strength of regularization is increased by a factor of $2$ to match the image resolution.

\begin{figure}[!hb]
  \includegraphics[width=1\columnwidth,height=5cm]{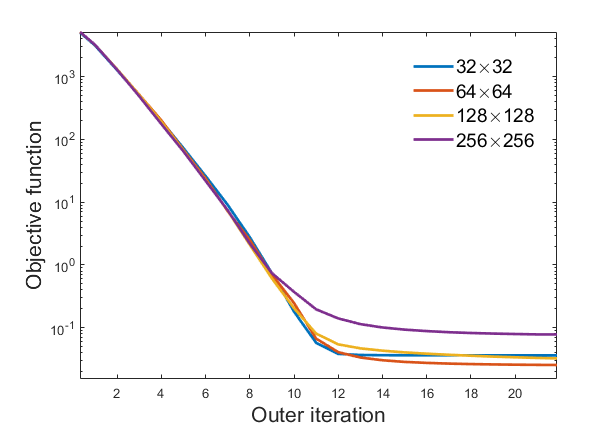}
  \caption{Objective function vs. number of outer iterations with different problem size}
  \label{Figure:size}
\end{figure}

The number of iterations required to converge is 14, 18, 19 and 20, respectively, as shown in Figure \ref{Figure:size}. This indicates that PN algorithm has almost perfect scalability with respect to problem size, similar to that of the classical Newton method for smooth problems.

\subsection{Effect of Hessian computation}

\begin{figure}[!hb]
  \includegraphics[width=1\columnwidth,height=5cm]{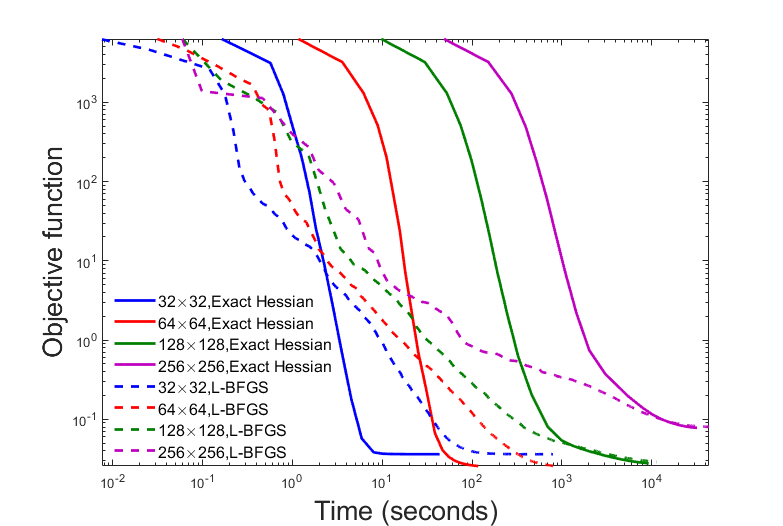}
  \caption{Objective function vs. time with different Hessian computation methods}
  \label{Figure:hessian}
\end{figure}

The BFGS algorithm approximates the Hessian matrix and its inverse from differences in the gradient and increments in the solution between iterates, and stores such approximation as a dense matrix. As previously discussed, storing the Hessian (or its approximation) as a dense matrix is unfeasible for medium to large scale problems, because of storage requirements. To address this issue, limited memory-BFGS computes the approximated action of the Hessian and its inverse using a limited number of updating vectors, thus drastically reducing storage requirements. Here, we compare the convergence of the PN algorithm using the exact Hessian and its L-BFGS approximation. In all experiments, we limit the maximum number of L-BFGS update vectors to 50. Results are shown in Figure \ref{Figure:hessian}. 

For relatively small problem sizes ($32\times32$), the L-BFGS approximations achieves better performance in the early iterations but  exact Hessian quickly catches up and allows to reach the target accuracy faster in terms of time-to-solution. As the problem dimension increases, L-BFGS approximation becomes more competitive due to the increased computational cost of evaluating the exact Hessian action. As shown in the figure, PN with L-BFGS approximation can outperform PN with exact Hessian with respect to time-to-solution for large-scale images and practical accuracy requirements.  

To sum up, if a coarse accuracy is needed, L-BFGS outperforms exact Hessian for large scale problems, however as one increases the accuracy requirements using the exact Hessian will eventually pay off. Another advantage of using the exact Hessian operator is that---when memory is the limiting factor---evaluating the action of the exact Hessian on a vector using Algorithm \ref{alg:2} only requires to store two temporary vectors with the same dimension of the data, while L-BFGS requires storing several updating vectors of the same dimension of the image. 


\section{Conclusion}

In this paper, we demonstrated an application of proximal Newton (PN) method to X-Ray tomography reconstruction problems with total variation regularization and synthetic sinogram data. At each iteration, the PN method minimizes a surrogate objective function where the smooth and computationally expensive data fidelity term is approximated by its second-order Taylor's expansion. 

Numerical results indicates that the PN algorithm can outperform state-of-the-art first-order proximal algorithms (such as FISTA) not only in terms of number of gradient evaluations but also in terms of time-to-solution for the application considered here. In addition, we numerically verified that the PN algorithm converges in a number of iterations that is almost independent of problem size, thus enjoying scalability properties similar to that of the classical Newton method for smooth problems. 

Finally, we compared the effect of different Hessian approximations on PN performance. In particular, the L-BFGS approximation of the Hessian drastically reduces the computational cost of minimizing the PN surrogate objective functions, thus allowing for faster (in terms of time) progress in the early iterations of the algorithm. This speed-up in the initial phase of the algorithm becomes more and more evident as the resolution is increased. However, when the optimization problem needs be solved with high accuracy, using the exact Hessian will eventually outperforms the L-BFGS approximation due the its higher rate of convergence.





\bibliographystyle{unsrt}
\bibliography{refs}


\begin{biography}
\textbf{Tao Ge} received his BS in electrical engineering from Harbin Engineering University (2016) and his master in electrical engineering from Washington University in St. Louis (2018). Since then he has been a PhD candidate in electrical engineering of Washington University in St. Louis. His work has focused on CT reconstruction problems.

\textbf{Umberto Villa} received his BS and MS in Mathematical Engineering from Politecnico of Milano (Milan, Italy, 2005 \& 2007) and his Ph.D. in Mathematics from Emory University (Atlanta, US, 2012). He joined the Electrical \& Systems Engineering Department of Washington University in St. Louis in 2018. His work has focused on the computational and mathematical aspects of large-scale inverse problems, medical imaging, uncertainty quantification, optimal experimental design. He is a member of IEEE and SIAM.

\textbf{Ulugbek S. Kamilov} obtained his BSc and MSc in Communication Systems, and PhD in Electrical Engineering from EPFL, Switzerland, in 2008, 2011, and 2015, respectively. Since 2017, he is an Assistant Professor and Director of Computational Imaging Group (CIG) at Washington University in St. Louis. His research area is computational imaging, large-scale optimization, and machine learning with applications to optical microscopy, magnetic resonance and tomographic imaging. In 2017, he received the IEEE Signal Processing Society’s Best Paper Award. He is currently a member of IEEE Technical Committee on Computational Imaging and Associate Editor for the IEEE Transactions on Computational Imaging.


\textbf{Joseph O'Sullivan} joined the faculty in the Department of Electrical Engineering at Washington University in St. Louis (WashU) in 1987. He was the Director of the Electronic Systems and Signals Research Laboratory (ESSRL) from 1998 to 2007. Today, Professor O'Sullivan directs the Imaging Science and Engineering certificate program.

\end{biography}

\end{document}